\documentclass{article}
\PassOptionsToPackage{numbers, compress}{natbib}

\usepackage[preprint]{neurips_2024}

\usepackage{pgfplots}
\usepackage[utf8]{inputenc} 
\usepackage[T1]{fontenc}    
\usepackage{hyperref}       
\usepackage{url}            
\usepackage{booktabs}       
\usepackage{amsfonts}       
\usepackage{nicefrac}       
\usepackage{microtype}      
\usepackage{xcolor}         
\usepackage{graphicx}
\usepackage{pifont}
\usepackage{makecell}
\usepackage{multirow}
\usepackage{multicol}
\usepackage{colortbl}  
\usepackage{amsmath}
\usepackage[capitalize]{cleveref}  

\pgfplotsset{compat=1.18}
\pdfoutput=1

\title{\includegraphics[scale=0.13]{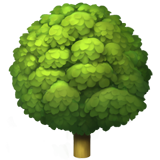} OpenCarbonEval: A Unified Carbon Emission Estimation Framework in Large-Scale AI Models}

\author{%
    Zhaojian Yu$^1$\thanks{~~Equal contribution.},
    Yinghao Wu$^{1*}$, 
    Zhuotao Deng$^1$,
    Yansong Tang$^{1\dagger}$,
    Xiao-Ping Zhang$^1$\thanks{~~Corresponding author.} \\
  $^1$Shenzhen Ubiquitous Data Enabling Key Lab, Tsinghua University\\
  \texttt{\{yzj23,yh-wu23,dzt23\}@mails.tsinghua.edu.cn} \\
  {\url{https://github.com/answers111/OpenCarbonEval}}
  \vspace{-0.5cm}
}

\begin{document}

\maketitle

\begin{abstract}
In recent years, large-scale auto-regressive models have made significant progress in various tasks, such as text or video generation. However, the environmental impact of these models has been largely overlooked, with a lack of assessment and analysis of their carbon footprint. 
To address this gap, we introduce OpenCarbonEval, a unified framework for integrating large-scale models across diverse modalities to predict carbon emissions, which could provide AI service providers and users with a means to estimate emissions beforehand and help mitigate the environmental pressure associated with these models. In OpenCarbonEval, we propose a dynamic throughput modeling approach that could capture workload and hardware fluctuations in the training process for more precise emissions estimates.
Our evaluation results demonstrate that OpenCarbonEval can more accurately predict training emissions than previous methods, and can be seamlessly applied to different modal tasks. Specifically, we show that OpenCarbonEval achieves superior performance in predicting carbon emissions for both visual models and language models. By promoting sustainable AI development and deployment, OpenCarbonEval can help reduce the environmental impact of large-scale models and contribute to a more environmentally responsible future for the AI community.

\end{abstract}

\begin{figure}[h]
  \includegraphics[width=0.95\textwidth]{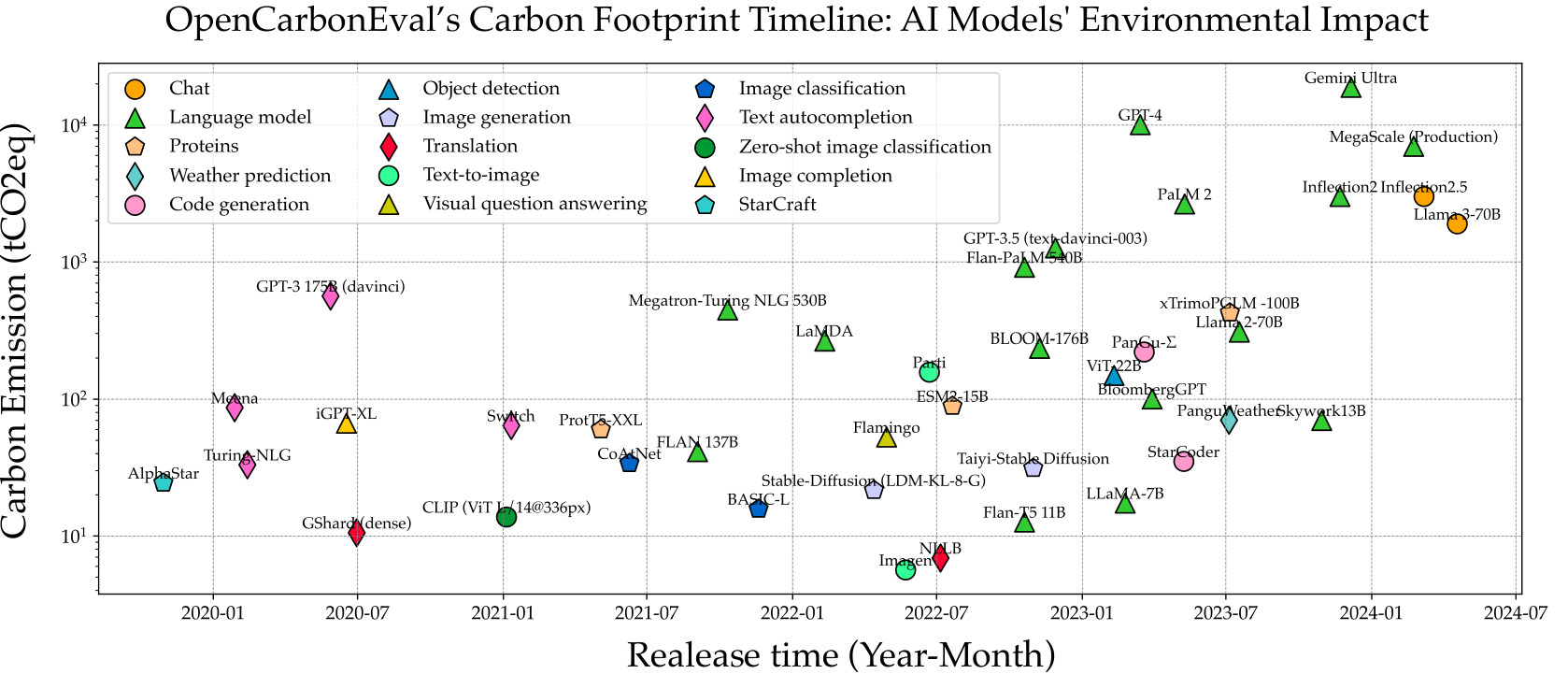}
  \caption{Large-scale models' environmental impact covering 42 large-scale AI models across 15  tasks. OpenCarbonEval enables the estimation of carbon emissions for various models, facilitating a more sustainable AI development process.}
  \label{fig:timeline}
\end{figure}

\section{Introduction}

Recently, transformer-based Large Language Models (LLMs) and Large Multimodal Models (LMMs) have exhibited remarkable intelligence across a wide range of tasks, largely attributed to the advancement of their scaling laws~\cite{henighan2020scaling,kaplan2020scaling,zhai2022scaling}. However, as the scale of model parameters and training sets increases, the computational overhead of training and maintaining large-scale models becomes substantial, resulting in significant environmental impacts.
For instance, the creation of GPT-3~\cite{brown2020language} with 175 billion parameters emitted approximately 552 tonnes of \textit{carbon dioxide equivalent} (CO2eq), which is three times the CO2eq emissions of jet plane round trip between San Francisco and New York~\cite{patterson2021carbon}.
Consequently, the serious ecological implications of large-scale models necessitate that AI service providers and users are cognizant of the carbon footprint of various emerging large-scale models.

Although various large-scale models have made significant progress~\cite{llama3modelcard,achiam2023gpt}, researches about their environmental impacts has lagged far behind. 
Previous work,  such as MLCO2~\cite{lacoste2019quantifying} and GreenAlgorithm~\cite{lannelongue2021green}, have proposed to calculated the carbon emission of Machine Learning (ML) tasks based on some key parameters like GPU usage, training duration, and data center efficiency. However, these methods are limited to small-scale models and have failed to keep pace with the rapid scaling of large-scale models.
In response to the rapid scaling of large-scale models, a growing body of research has shifted its focus to their carbon footprint, driven by the significant increase in energy consumption. 
For example, BLOOM's carbon footprint report~\cite{JMLR:v24:23-0069} proposes a Life Cycle Assessment (LCA) estimation approach, tracing the operational carbon and embodied carbon generated by BLOOM~\cite{workshop2022bloom} throughout its entire life cycle. 
However, a key limitation of LCA lies in its retrospective nature, only allowing for the estimation of emissions after model training, rather than providing foresight into potential emissions before training begins. 
To break away from this limitation, LLMCarbon~\cite{faiz2023llmcarbon} presents an end-to-end approach for carbon emission predicting, leveraging training configuration information from preceding models as features in a polynomial regression analysis to derive average hardware efficiency across diverse hardware setups, and subsequently estimates potential emissions based on the regression-derived efficiency. However, the regression-based carbon emission prediction method is limited to considering only static features in model training, which can lead to biased predictions that are heavily influenced by the training data and methods. Moreover, the modality-specific design of current estimation methods hinders the development of a unified framework that can seamlessly accommodate diverse modalities.

In this paper, we propose OpenCarbonEval, a unified framework integrating large-scale models across diverse modalities, which could accurately predict the potential carbon emission before model training.
OpenCarbonEval integrates Little's Law~\cite{little2008little} with a novel Dynamic Throughput Modeling method to adapt to changing computational workloads and hardware configurations, 
which enables a more comprehensive and reliable carbon emission prediction process by moving beyond simplistic polynomial regression tasks. 
To validate our approach, we evaluate 7 various large-scale models covering vision and language task and compare the predicted emissions with their published carbon footprint.
Experimental results demonstrate that OpenCarbonEval successfully alleviates the uncertainty of carbon emission predictions, yielding a substantial reduction in prediction errors. By leveraging OpenCarbonEval, we conduct an in-depth analysis of large-scale carbon emissions, thereby providing valuable insights into mitigating the environmental footprint of large-scale models.

\section{OpenCarbonEval}

\begin{figure}[t]
  \centering
  \includegraphics[width=\textwidth]{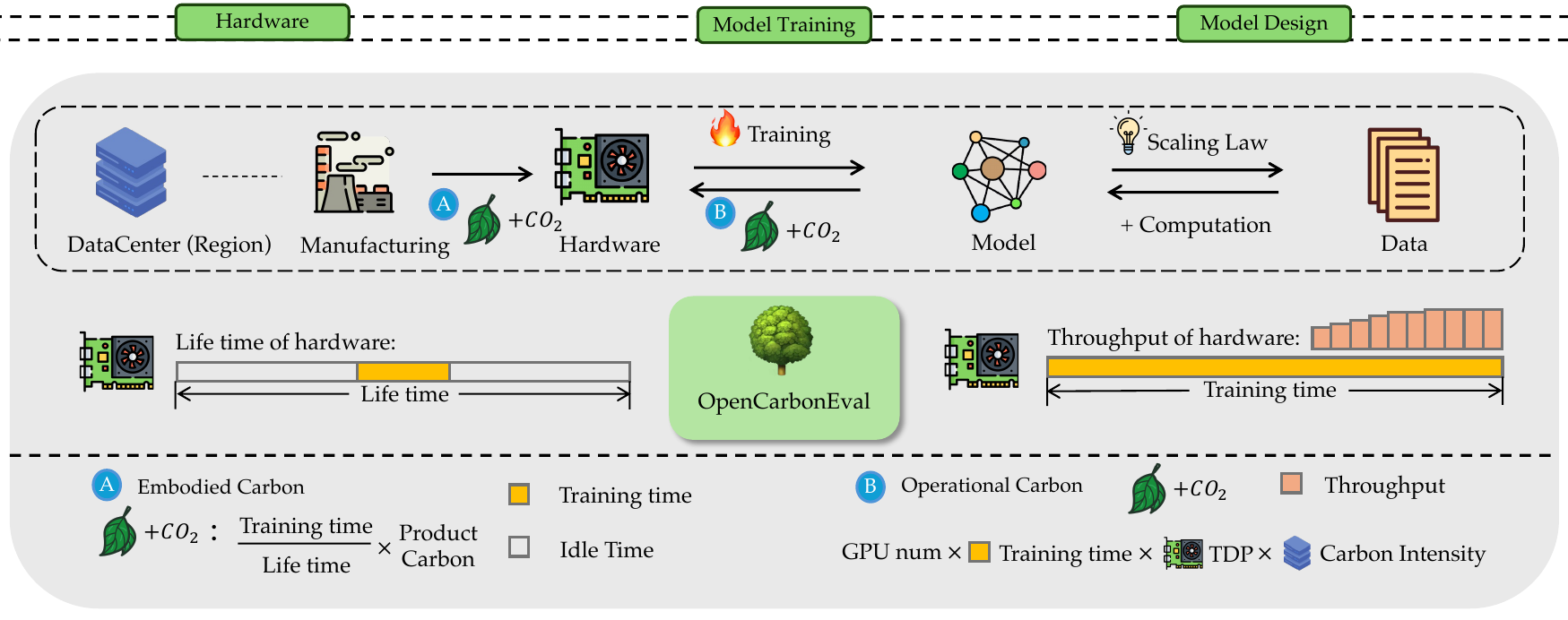}
  \caption{The overall pipeline of OpenCarbonEval. OpenCarbonEval leverages Little's Law to establish a rigorous framework for modeling computation, throughput, and time, thereby enabling the precise estimation of carbon emissions associated with the model.}
  \label{opencarboneval}
\end{figure}

Our investigation begins with a critical recap on how large-scale model produces carbon emission during its lifelong cycle. As shown in Figure \ref{opencarboneval}, the carbon emission of large-scale models mainly includes two types: Operational Carbon and Embodied Carbon. In this section, we delineate the estimation methodologies for both types of carbon emissions, which subsequently inform the development of our Dynamic Throughput Modeling framework.

\subsection{Operational Carbon}
Operational carbon, produced by generating the electricity necessary for powering model training, is often calculated by multiplying the number of GPU hours used by the thermal design power (TDP) of those GPUs and the carbon intensity ($I$) of the energy grid used to power the hardware, which can be written as follows:
\begin{equation}
\label{base}
    C = E \cdot I =  P \cdot \mathcal{T} \cdot I
\end{equation}
where $C$ indicates the amount of emitted carbon dioxide ($kgCO2eq$), $I (kgCO_2eq/KWh)$ indicates the emitted $CO_2$ per $KWh$ energy consumed, $E$ indicates the energy consumed for model training ($KWh$), and $P$ is often set by 
Thermal Design Power (TDP) due to the unavailability of real-time power consumption and $\mathcal{T} (GPUh)$ represents the GPU-time product. 

\subsection{Embodied Carbon}
Embodied carbon, representing the emission associated with hardware manufacturing and the processes involved in producing a given product, is calculated as follows:
\begin{equation}
\label{embodied}
    C' =\sum_{r=0}^n\frac{T}{T'_r}\cdot C_r
\end{equation}
where $C'$ and $T$ indicate the embodied carbon and training time of the model to be estimated respectively, $C_r$ and $T'_r$ represent the product carbon and life time of the $r$-th GPU respectively, $n$ is the number of all hardware involved in training process.  

\subsection{Little's Law}
In Equation \ref{base}, the grid's carbon intensity $I$ is a coefficient ($kgCO2eq/kWh$) depends on the electricity source that powers training process which is often related to the region where the data center is located and $P$ is often a certain value given the GPU type. Therefore, the accuracy of operational carbon emission estimating prior to model training hinges decisively on obtaining an extremely precise measurement of GPU-time product. However, to this day, the precise measurement of GPU-time product still suffers from the misalignment of different model architecture, different accelerating methods and different hardware efficiency.
To solve this problem, we draw inspiration from queuing theory and use Little's Law \cite{little2008little} to model the relationship between total computation, training speed and GPU time during the model training process.  
\begin{equation}
\label{ll}
    Concurrency=Latency \times Throughput
\end{equation}
Little's Law, which can be written as the above formula, points at that the number of processed item in a system (Concurrency) is equal to the multiplication of processing speed (Throughput) and the average processing time (Latency). 
In our approach, we let $T$ indicates the total training time and divide $T$ into the same $n$ parts. Given a time $t\in[0,T]$, the training speed  at time $t$ can be considered as $f(t)$. When $n$ is very large, we can get the following results by applying Little's law in a short period of time:

\begin{equation}
\label{sum}
    \mathcal{C}(P, D) = \lim_{n\to\infty}\frac{T}{n}\sum_{k=0}^{n}f\left(\frac{k\cdot T}{n}\right)=\int_{0}^{T}f(t)dt
\end{equation}

where $\mathcal{C}$ indicates the Computation (Concurrency) in the training process, which is a certain value that can be accurately calculated by the model parameter ($P$) and training data size ($D$). From Equation \ref{sum}, we can solve for the training time $T$ and bring it into Equation \ref{base} to obtain operational carbon.

\subsection{Dynamic Throughput Modeling}
\begin{figure}[t]

  \centering
  \includegraphics[width=\textwidth]{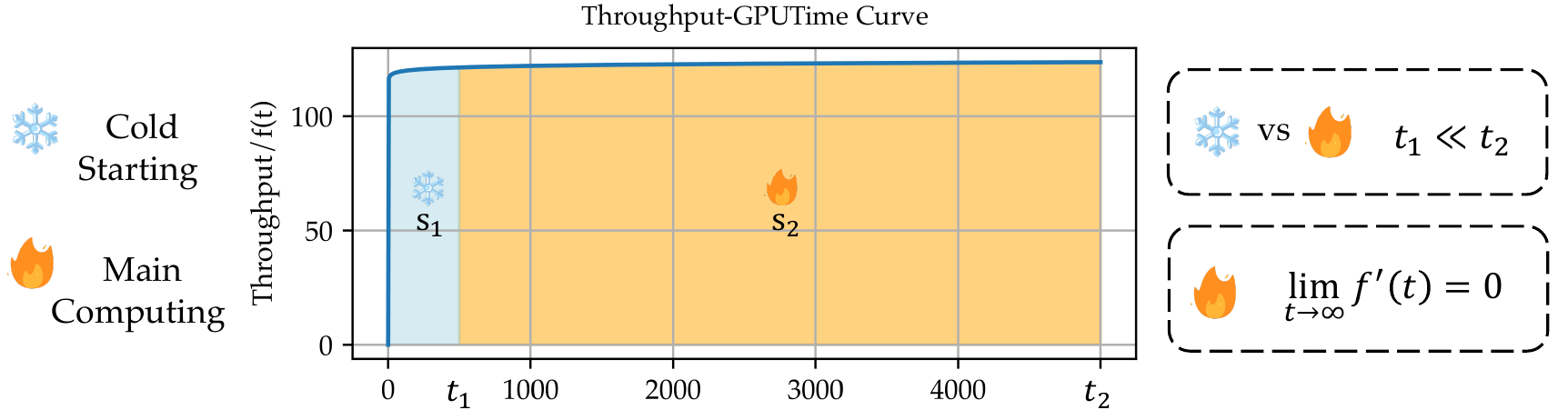}
  \caption{The two-stage modeling about hardware performance. The cold starting phase is characterized by a brief duration of $t_1$, after which the actual training commences. Subsequently, the entire training process is completed within a total time span of $t_2$.}
  \label{curve}
\end{figure}

As illustrated in Equation \ref{sum}, the training time $T$ could be calculated given the total computation and throughput $f(t)$. However, the throughput $T$ is often difficult to reach due to different hardware configurations and training setup.
To this end, different from the previous work~\cite{faiz2023llmcarbon} who use a polynomial regression model to predict the hardware performance directly, we hope OpenCarbonEval can obtain the hardware performance as accurately as possible. Therefore,  we conduct a two-stage dynamic modeling of hardware performance and obtain more accurate throughput results. 

As illustrated in Figure \ref{curve}, we divide the entire training process into two stages: the cold start stage and the main computing stage. The cold start stage includes the loading process of the model and data and some preliminary computation, which often only accounts for a small part of the entire training process~\cite{narayanan2021efficient}.
Therefore, we can get that the cold start time $t_1$ is much smaller than the main computing time $t_2$. In addition, during the main computing phase after a cold start, the computing performance of the GPU tends to stabilize. Therefore, the Throughput-GPUTime curve $f(t)$ often needs to satisfy that $\lim_{t\to\infty}f'(t)=0$. 
To meet the above two requirements, in our experiments, we use the logarithmic function $f(t)=ln(1+\alpha t)$ to model the hardware throughput where $f(t)$ indicates the training speed (the throughput in Equation \ref{ll}). Unless otherwise specified, we use throughput-$\alpha$ to refer to $\alpha$ in $f(t)$ in this paper.

By leveraging dynamic throughput modeling, OpenCarbonEval predicts carbon emissions through a three-stage process: 1) computing required Computation (FLOPs) based on model architecture, parameters, and training data scale, 2) modeling Throughput-$\alpha$ specific to the device type, 3) calculating precise emissions using Equation \ref{base} and \ref{embodied}.

\section{Experiments}

\subsection{Setup}
\textbf{Baselines} In this section, we present an empirical evaluation that demonstrates the superiority of OpenCarbonEval and Dynamic Throughput Modeling (DTM) in carbon emission prediction, showcasing its performance advantages over the following regression-based static throughput modeling approaches: polynomial regression, the method used by LLMcarbon, support vector regression (SVR) and decision tree regression (DTR). 

\textbf{Dataset} 
In our experimental setup, we leverage real-world data points sourced from EpochAI~\cite{epoch2023aitrends}, which provide a rich repository of model training details, including computation requirements and hardware specifications, thereby enabling a more realistic and informed evaluation of our approach.
Moreover, to ensure a fair comparison, we curated a diverse set of open-source large-scale models, varying in functionality, input data, geographical region, and computing equipment used for training to serve as test data points. 
We present results from an array of open-sourced LLMs, such as ChatGLM~\cite{zeng2022glm}, a bilingual (English and Chinese) pre-trained language model with 130 billion parameters, StarCoder~\cite{li2023starcoder}, a generative model for code synthesis and LLaMa-3-70B~\cite{llama3modelcard}, a model trained on Meta’s large-scale AI clusters which takes data and scale to new heights. While the scaling laws of language models have been well-established, those of visual models remain an active area of exploration, with a notable absence of carbon emission predictions for this type of model. Consequently, we selected two iconic models, Vision Transformer (ViT)~\cite{dosovitskiy2020image} and Swin Transformer~\cite{liu2021swin}, to incorporate into our experiments.
\begin{table}[t]
\caption{Throughput-$\alpha$ of different devices. }
\label{tab:interval}
\centering
\begin{tabular}{l|cc|c}
\toprule
\centering
\textbf{Device} & \textbf{Thermal Design Power (W)}     &\textbf{Peak TFLOPs/s} & \textbf{Throughput-$\alpha$} \\
\midrule
Google TPUv2     &280& 46           & $10^{-1}\sim10^{-0.5}$                       \\
Google TPUv3     &450& 123           & $10^{3}\sim10^{13}$                       \\
Google TPUv4     &300& 275           & $10^{15}\sim10^{40}$                       \\
\midrule
Nvidia Tesla K80 & 300& 8.73           & $10^{-6.7}\sim10^{-6.6}$                         \\
Nvidia Tesla P100& 300& 21.2           & $10^{-4.5}\sim10^{-3.8}$                         \\
Nvidia Tesla V100&300& 125           & $10^{-2}\sim10^{10}$                         \\
Nvidia Tesla A100 &400& 312           & $10^{20}\sim10^{50}$                    \\
Nvidia Tesla H100 &700& 989         & $10^{100}\sim10^{110}$                    \\
\bottomrule
\end{tabular}
\end{table}
\begin{figure}[tp]

    \includegraphics[width=\textwidth]{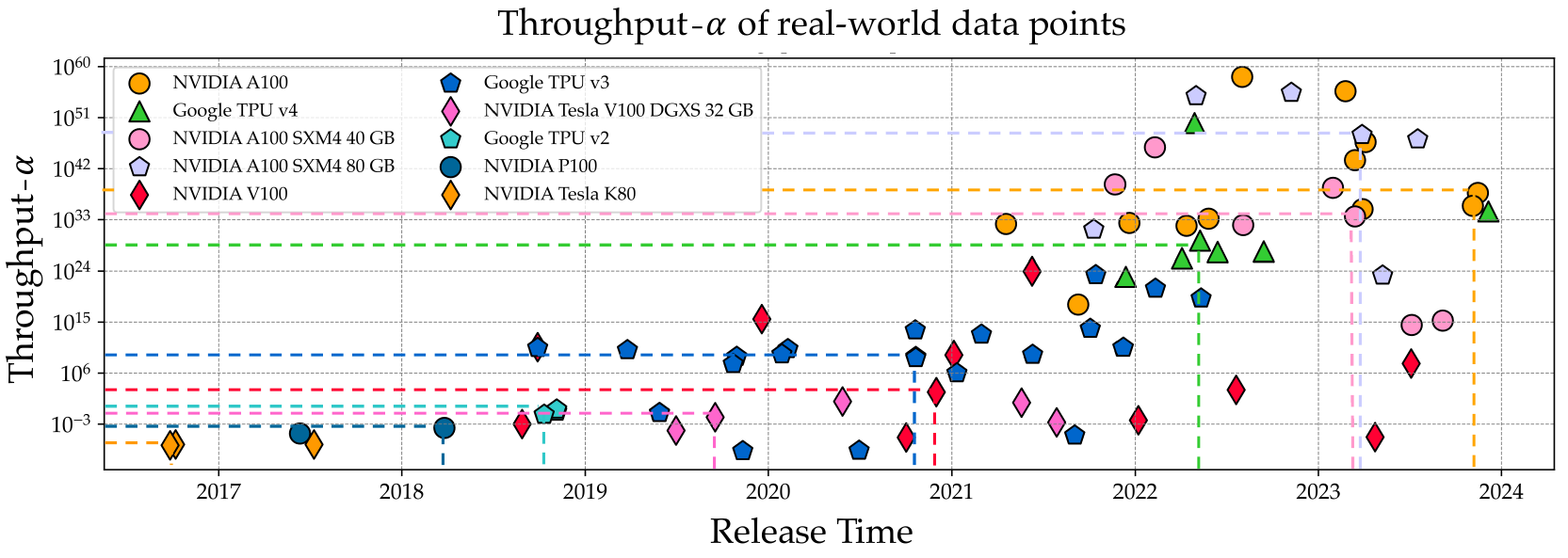}
    \caption{The data points indicate the model under different hardware. In our experimental setup, we aggregated hardware devices sharing a common prefix, such as Nvidia A100, into a single category to facilitate analysis and comparison.}
    \label{fig:alpha}
\end{figure}

\subsection{The Equipment Era Effect Law}

\begin{figure}[tp]

    \includegraphics[width=\textwidth]{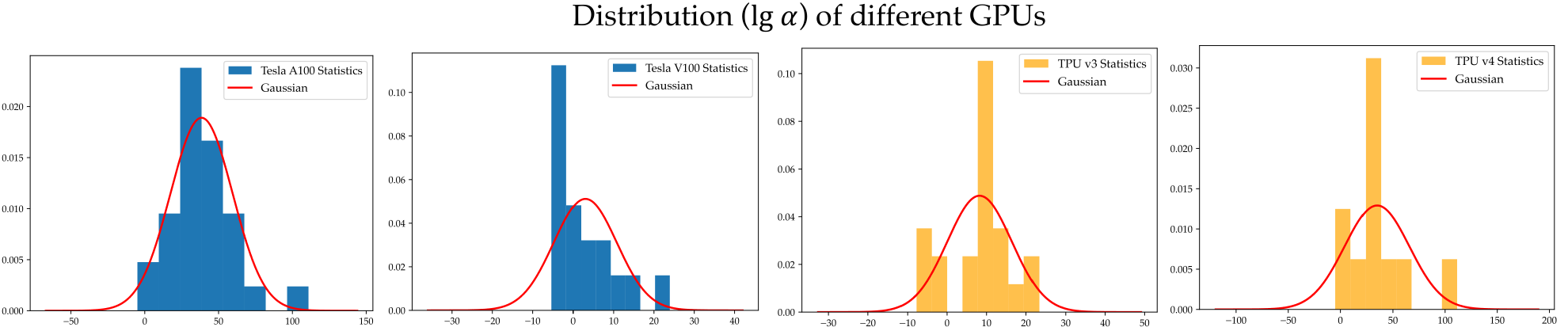}
    \caption{Distribution about different hardware. The throughput-$\alpha$ distribution is centered around the mean, exhibiting a pronounced clustering of values in close proximity to the average, thereby underscoring the stability of our model's performance.}
    \label{fig:gauss}
\end{figure}

Figure \ref{fig:alpha}, \ref{fig:gauss} present the throughput-$\alpha$ values we obtained for different models and different devices, highlighting the following salient observations:

\textbf{The range of throughput-$\alpha$ is predominantly determined by the generation of training equipment used.} It is important to acknowledge that while throughput-$\alpha$ values for the same device may vary, these variations are typically not orders of magnitude apart. This is because the underlying hardware architecture and computational capabilities of the device remain consistent. However, when comparing different devices, such as GPUs from different generations or manufacturers, the throughput-$\alpha$ values can exhibit more significant discrepancies due to differences in their design, performance, and optimization techniques.

\textbf{Within the same device pool, the choice of model architecture and hyperparameters can lead to throughput-$\alpha$ gaps.} This is because different models have varying computational requirements, memory access patterns, and levels of parallelism. For instance, a complex model with a large number of parameters and intricate computations may exhibit lower throughput-$\alpha$ compared to a simpler model that is more efficient in utilizing the device's resources. Additionally, the choice of optimization algorithms, batch size, and other training configurations can also impact the throughput-$\alpha$ of a model.

As illustrated in Figure \ref{fig:gauss}, despite efforts to normalize the data, some variance remains in the distribution of data points. To mitigate the impact of this variance on our predictions, we adopt a conservative approach by reporting the interval $(\mu-0.6\sigma,\mu+0.6\sigma)$ as our predicted range of potential carbon emissions, which is shown in Table \ref{tab:interval}. For the purpose of benchmarking against other methods in OpenCarbonEval, we utilize the midpoint of this interval as a representative value.

\begin{table}[t]
\centering
\renewcommand{\arraystretch}{1.0}
\caption{Operational carbon of various LLMs on different GPU. The result of the best method is bolded. $\Delta$ represents the relative error between the predicted value and the actual value.}
\label{table:alpha_t}
\setlength{\tabcolsep}{1.9mm}{
\begin{tabular}{l|cccc|cc}
\toprule
\textbf{Method}                                                                   & \textbf{GLM}    & \textbf{BLOOM}      & \textbf{StarCoder}  & \textbf{LLaMa-3 }   & \textbf{ViT-L/16} & \textbf{Swin-L}    \\
\midrule
Params                                                                    & 130B       & 176B       & 15B        & 70B        & 307M     & 197M       \\
ZettaFLOPs        & 312        & 387        & 93         & 6300       & 0.53     & 0.40       \\
Device                                                             & A100 & A100 & A100 & H100 & TPUv3    & V100 \\

$I$ ($gCO2eq/kWh$)  & 581 & 57 &155  & 424 & 369 & 369 \\
\midrule
Actual CO2eq ($t$)                                                           &       257     &         24.7   &     17.26       &   1900         &    2.71      &     0.80    \\
\midrule
\multicolumn{7}{c}{Static Modeling}\\
\midrule
LLMCarbon                                                          &   153.11  &    19.89   &    14.14    &    4074.63    &  0.20    &    0.10              \\
$\Delta$                                                           &      -40.4\%      &     -19.4\%       &      -18.1\%      &     +114.5\%       &     -92.6\%       &          -87.5\%            \\
\midrule
SVR                                                                 & 160.21  & 21.49 & 14.01 & 811.20 & 0.45 & 0.25 \\
$\Delta$                                                           &    -37.7\%        &    -13.0\%   &      -18.8\%      &     -57.31\%       &    -83.3\%        &         -68.8\%              \\
\midrule
DTR                                                                 & 161.39  & 21.42 & 14.32 & 803.64 & 1.03 & 0.51  \\
$\Delta$                                                           &    -37.2\%        &      -13.2\%      &      -16.7\%      &      -57.70\%       &     -62.0\%       &        -36.3\%            \\
\midrule
\multicolumn{7}{c}{Dynamic Modeling}\\
\midrule
OpenCarbonEval                                                     &    \textbf{276.92}   &    \textbf{21.96}   &    \textbf{18.07}    &    \textbf{1966.17}   &       \textbf{2.29}  &    \textbf{0.68}              \\
$\Delta$                                                           &  \textbf{+7.8\%}          &      \textbf{-11.1\%}       &      \textbf{+4.7\%}      &     \textbf{+3.5\% }      &    \textbf{+15.4\% }       &  \textbf{-14.5}\%              
\\
\bottomrule
\end{tabular}
}
\end{table}

\subsection{Operational Carbon}
Table \ref{table:alpha_t} presents the result of OpenCarbonEval on various large-scale models. We have compiled a comprehensive table that outlines all the parameters necessary for predicting carbon emissions. Within this table, ZettaFLOPs represent the total computation amount required for effective model training. For the actual CO2eq values, we rely on self-reported data where available. Otherwise, we employ other relevant real data provided and obtain it according to Equation \ref{base}.

\textbf{Compared with static modeling, OpenCarbonEval exhibits a significantly lower relative error in predicting carbon emissions.}
In contrast to the actual CO2eq emissions, the static modeling methods and LLMCarbon exhibited significant errors, with a notable discrepancy of up to 114.5\% in predicting the training carbon footprint. This is attributed to the inability of static modeling to capture the dynamic law of throughput. In stark contrast, our proposed method, opencarboneval, demonstrated remarkable accuracy, with small relative errors at all test data points, thereby validating its effectiveness.

\textbf{OpenCarbonEval consistently achieves low relative errors in its predictions for both visual and language models, demonstrating its versatility and robustness across different modalities.}
Notably, when predicting the carbon footprint of visual models such as ViT/16-L and Swin-L, OpenCarbonEval still outperforms LLMCarbon, achieving relatively accurate predictions. This superiority can be attributed to OpenCarbonEval's unique strength in establishing a unified task set that can accommodate all modalities, which is facilitated by its computation-throughput-latency ternary architecture that incorporates dynamic modeling for throughput.
\subsection{Embodied Carbon}

For embodied carbon estimation, we reformulated Equation \ref{embodied} as follows:
\begin{equation}
    C' = \mathcal{T} \cdot \beta
\end{equation}
where $\beta$ indicates the $CO_2$ emitted per GPUh in the life time of given GPU, $\mathcal{T}$ represents the GPU-time product.
By reviewing LLMCarbon and obtaining specifications for different types of hardware materials, we calculated the carbon emissions per $gpuh$, assuming a 1-year effective lifespan for each hardware component. This approach allows us to account for the embodied carbon emissions resulting from the manufacturing process, which is an essential aspect of comprehensive carbon evaluation. As shown in Table \ref{tab:embodied}, although embedded carbon constitutes a relatively small proportion of the total carbon evaluation, OpenCarbonEval can still maintain high prediction accuracy, demonstrating the effectiveness of our approach in capturing the nuances of carbon emissions in AI model training.

\begin{table}[t]
\caption{Different embodied carbon prediction results on various models by OpenCarbonEval.}
\label{tab:embodied}
\centering
\label{table:embodied}
\setlength{\tabcolsep}{1.8mm}{
\begin{tabular}{l|cccc|cc}
\toprule
~ & \textbf{GLM}    & \textbf{BLOOM}      & \textbf{StarCoder}  & \textbf{LLaMa-3 }   & \textbf{ViT-L/16} & \textbf{Swin-L}    \\
\midrule
Device                                                             & A100 & A100 & A100 & H100 & TPUv3    & V100 \\


TSMC process  &    7 nm   &    7 nm   &    7 nm       &   4 nm         &    16 nm      &     12 nm   \\
Die size  &     826 $mm^2$     &        826 $mm^2$   &     826 $mm^2$       &   814 $mm^2$         &    700 $mm^2$     &     815 $mm^2$   \\
\textbf{$gCO2eq/GPUh$}  &     1.5     &        1.5   &    1.5     &   1.7        &    0.8     &  1.1  \\
\midrule
Actual embodied &     \multirow{2}{*}{1634.50}     &         \multirow{2}{*}{1631.23}   &     \multirow{2}{*}{480.38}       &   \multirow{2}{*}{10880.0}         &    \multirow{2}{*}{13.06}      &     \multirow{2}{*}{7.92}   \\  \multicolumn{1}{c|}{$CO_2$eq (kg)} 
&     &         &     &   &    &     \\

\midrule
LLMCarbon                                                          &    898.37   &    1090.65    &    285.23    &  21211.80 &  0.88  &    0.91              \\
$\Delta$                                                           &      -45.0\%      &     -33.1\%       &      -40.6\%      &     +95.0\%       &     -93.3\%       &          -88.5\%            \\
\midrule
OpenCarbonEval  &    \textbf{1787.35}   &    \textbf{1444.75}   &    \textbf{437.11}    &    \textbf{11261.75}   &       \textbf{11.05}  &    \textbf{6.77}              \\
 $\Delta$                                                           &  \textbf{+9.4\%}          &      \textbf{-11.4\%}       &      \textbf{-9.0\%}      &     \textbf{+3.5\% }      &    \textbf{-15.4\% }       &  \textbf{-14.5}\%              
\\
\bottomrule
\end{tabular}}
\end{table}

\section{Analysis}
\begin{figure}[t]
    \centering
    
    \includegraphics[width=0.93\textwidth]{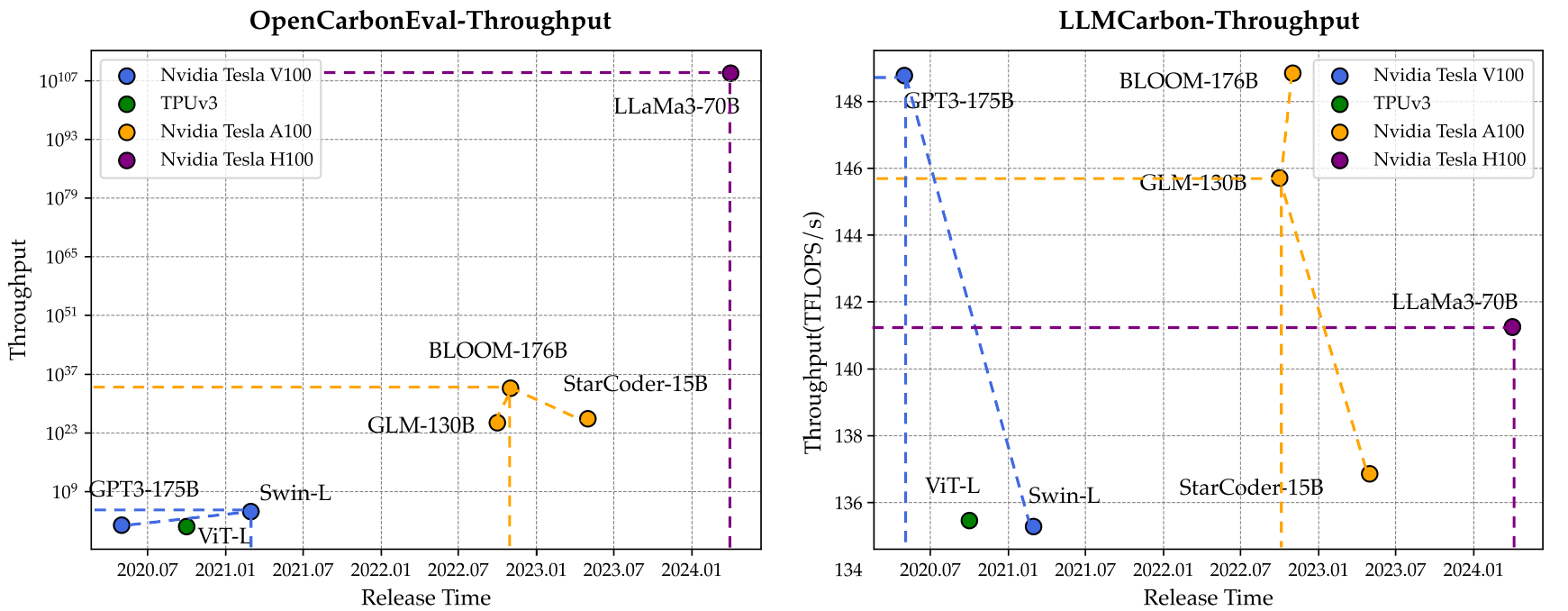}
    \caption{Throughput comparison between OpenCarbonEval and LLMCarbon}
    \label{fig:TM}
\end{figure}

\subsection{Throughput Modeling}
Figure \ref{fig:TM} presents a comparative analysis of OpenCarbonEval and LLMCarbon in predicting throughput, a critical factor in estimating carbon emissions. Specifically, we employ throughput-$\alpha$ as a proxy to quantify the computing processing speed of OpenCarbonEval, whereas LLMCarbon's predicted actual FLOPs per second are used as a metric to evaluate its performance. The trend revealed in Figure \ref{fig:TM} demonstrates that OpenCarbonEval adheres to the The Equipment Era Effect Law, which suggests that the throughput-$\alpha$ values for the same device remain relatively consistent across different workloads. In contrast, LLMCarbon struggles to accurately reflect the underlying hardware architecture, leading to discrepancies in its throughput predictions. Therefore, the intergenerational gap in hardware capabilities, which arises from differences in design, performance, and optimization techniques across various generations of devices, leads to erroneous throughput predictions. This, in turn, constitutes the primary source of error in forecasting carbon emissions, highlighting the need for more accurate and hardware-aware models.

\subsection{Carbon Footprint Scaling}
\begin{figure}[t]
    \centering
    \includegraphics[width=0.93\textwidth]{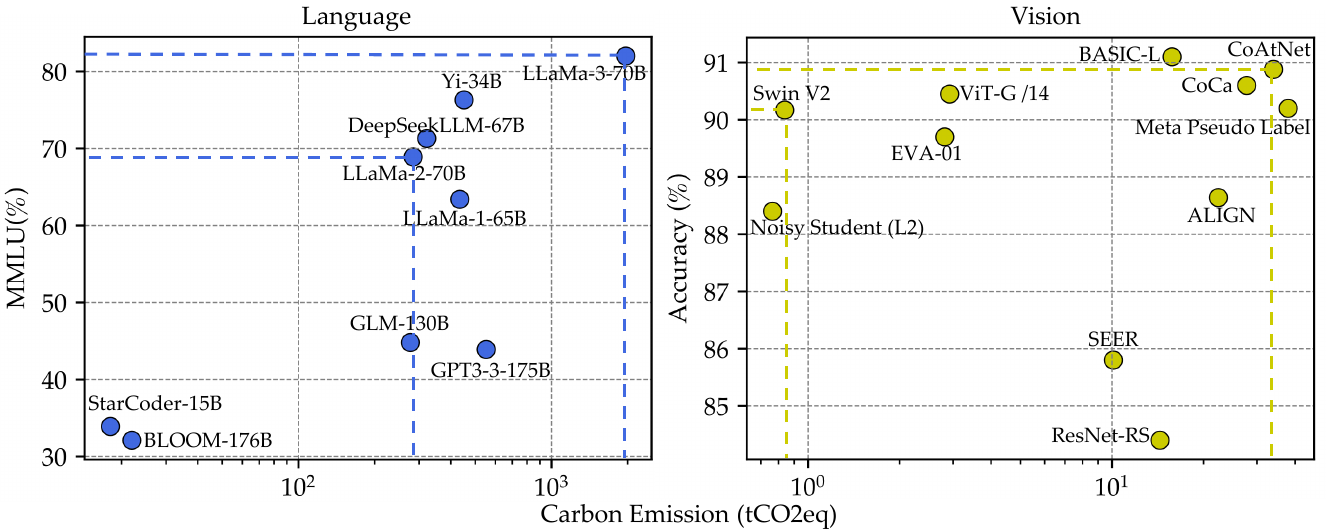}
    \caption{The trade-off between training carbon footprint and performance about different models}
    \label{fig:trade}
\end{figure}

We employ OpenCarbonEval to investigate the carbon footprint scaling of various visual and language models, with MMLU~\cite{hendryckstest2021} and ImageNet-1k Accuracy~\cite{deng2009imagenet} serving as performance indicators for language and vision models, respectively. As shown in Figure \ref{fig:trade}, our analysis reveals a notable trend: while language models exhibit a direct correlation between performance improvement and increased carbon emissions, likely due to the added computational costs associated with scaling up their intelligence, vision models do not always follow this pattern. In fact, certain models, such as the Swin Transformer, demonstrate improved computing efficiency by effectively harnessing the flexibility of deep networks. This discrepancy highlights the importance of considering the environmental impact of AI models and underscores the need for developers to strike a balance between performance improvement and efficiency optimization to minimize their carbon footprint.

\section{Related Work}

By analyzing the whole life cycle of a machine learning system, SustainableAI~\cite{wu2022sustainable} divides the industry-scale training and deployment carbon footprint of AI models into two parts: operational carbon and embodied carbon. Operational carbon includes the carbon emissions generated by producing the electricity required for training an AI model and using the model for inference in the location of the server. And embodied carbon means the equivalent carbon emissions from manufacturing the computing devices of the server.

To the best of our knowledge, the existing carbon emission estimation tools mainly focus on the evaluation of operational carbon footprint, which can be calculated by multiplying the energy required for AI computing by the regional carbon intensity $I$ ($kgCO_2$/$kWh$). 
To estimate the energy consumption of computing devices, MLCO2~\cite{lacoste2019quantifying} only calculates the GPU energy consumption by just simply multiplying the empirical average power corresponding to the given device type by the given computing time. Cumulator~\cite{trebaol2020cumulator} estimates the consumption of both CPU and GPU in a way similar to MLCO2~\cite{lacoste2019quantifying}, but it also lack of accuracy because it uses thermal design power (TDP) to replace the actual power during calculation, resulting in overestimation. Green-Algorithms\cite{lannelongue2021green}, CodeCarbon~\cite{lottick2019energy}, Eco2AI~\cite{budennyy2022eco2ai}, CarbonTracker~\cite{anthony2020carbontracker}, and EIT~\cite{henderson2020towards} have taken the energy consumption of both CPU, GPU and memory into consideration. Among them, CodeCarbon~\cite{lottick2019energy}, Eco2AI~\cite{budennyy2022eco2ai}, CarbonTracker~\cite{anthony2020carbontracker}, and EIT~\cite{henderson2020towards} uses software integrated tools (psutil and RAPL files) and internal tools (pynvml and nvidia-smi, only available for Nvidia GPUs) to respectively acquire the real-time CPU and GPU power, while Green-Algorithms uses fixed CPU and GPU power instead after considering the hardware TDP and utilization. 
LLMCarbon elaborates a predictive estimation framework for both dense and mixture-of-experts (MoE) LLMs, which contains a hardware efficiency model that allows us to predict the training time and then the operational carbon with arbitrary TPU/GPU numbers. Since we typically deploy AI models in data centers, most of these tools (apart from MLCO2~\cite{lacoste2019quantifying} and Cumulator~\cite{trebaol2020cumulator}) also consider additional overhead for cooling and infrastructure in the data center when computing energy consumption, The ratio of the total energy consumption to the energy consumption of computing devices is regarded as a fixed value related to the selected data center and defined as power usage effectiveness (PUE).

As for the embodied carbon, SustainableAI~\cite{wu2022sustainable} and LLMCarbon~\cite{faiz2023llmcarbon} have shown us how to calculate it. We can first obtain the energy consumption data of manufacturing each computing device from the manufacturer. Then the proportion of embodied carbon in manufacturing energy consumption is the same as the proportion of time required for AI tasks (training and/or inference) in the entire device lifetime.

\section{Discussion and Limitations}

\textbf{Lifecycle Emissions} While OpenCarbonEval provides a unified framework for estimating carbon emissions, it is not without its limitations. Notably, the current implementation does not account for the environmental impact of model deployment and inference, including data center operations, network transmission, and device usage. Future research directions could focus on expanding the framework to address these gaps and provide a more comprehensive understanding of AI's environmental footprint.

\textbf{Efficient Design} Another promising research direction is exploring carbon-efficient AI model design and optimization techniques, such as model pruning, knowledge distillation, and efficient neural network architectures, to reduce the carbon footprint of AI models without compromising performance. 

\textbf{Broder Impact on Environmental Sustainability} The increasing carbon footprint of large-scale AI models has significant implications for the environment and sustainability. Our analysis using OpenCarbonEval reveals a concerning trend of growing carbon emissions associated with the development and deployment of these models. This highlights the need for the AI community to prioritize environmental sustainability alongside performance and efficiency. Furthermore, the environmental impact of AI models can have far-reaching consequences, including contributing to climate change, air pollution, and e-waste generation.
By providing a unified framework for predicting carbon emissions, OpenCarbonEval can facilitate the development of more environmentally friendly AI models and encourage responsible AI practices. This includes promoting transparency and accountability in AI development, encouraging sustainable AI design and deployment, and fostering a culture of environmental responsibility within the AI community.
\section{Conclusion}
In this paper, we introduce OpenCarbonEval, a unified framework for integrating large-scale models across diverse modalities to predict carbon emissions. OpenCarbonEval is able to predict the training carbon emission of various large-scale AI models, resulting in a more carbon-transparent training process. Leveraging OpenCarbonEval, we conducted an analysis of the alarming growth of carbon emissions attributed to large models in recent years. Therefore, we call for every AI provider and user to pay more attention to the environment impact of large models and adopt more sustainable practices in their development and deployment.
{
\small
\bibliographystyle{unsrt}
\bibliography{main}

\begin{thebibliography}{10}

\bibitem{henighan2020scaling}
Tom Henighan, Jared Kaplan, Mor Katz, Mark Chen, Christopher Hesse, Jacob Jackson, Heewoo Jun, Tom~B Brown, Prafulla Dhariwal, Scott Gray, et~al.
\newblock Scaling laws for autoregressive generative modeling.
\newblock {\em arXiv preprint arXiv:2010.14701}, 2020.

\bibitem{kaplan2020scaling}
Jared Kaplan, Sam McCandlish, Tom Henighan, Tom~B Brown, Benjamin Chess, Rewon Child, Scott Gray, Alec Radford, Jeffrey Wu, and Dario Amodei.
\newblock Scaling laws for neural language models.
\newblock {\em arXiv preprint arXiv:2001.08361}, 2020.

\bibitem{zhai2022scaling}
Xiaohua Zhai, Alexander Kolesnikov, Neil Houlsby, and Lucas Beyer.
\newblock Scaling vision transformers.
\newblock In {\em Proceedings of the IEEE/CVF conference on computer vision and pattern recognition}, pages 12104--12113, 2022.

\bibitem{brown2020language}
Tom Brown, Benjamin Mann, Nick Ryder, Melanie Subbiah, Jared~D Kaplan, Prafulla Dhariwal, Arvind Neelakantan, Pranav Shyam, Girish Sastry, Amanda Askell, et~al.
\newblock Language models are few-shot learners.
\newblock {\em Advances in neural information processing systems}, 33:1877--1901, 2020.

\bibitem{patterson2021carbon}
David Patterson, Joseph Gonzalez, Quoc Le, Chen Liang, Lluis-Miquel Munguia, Daniel Rothchild, David So, Maud Texier, and Jeff Dean.
\newblock Carbon emissions and large neural network training.
\newblock {\em arXiv preprint arXiv:2104.10350}, 2021.

\bibitem{llama3modelcard}
AI@Meta.
\newblock Llama 3 model card.
\newblock {\em https://llama.meta.com/llama3}, 2024.

\bibitem{achiam2023gpt}
Josh Achiam, Steven Adler, Sandhini Agarwal, Lama Ahmad, Ilge Akkaya, Florencia~Leoni Aleman, Diogo Almeida, Janko Altenschmidt, Sam Altman, Shyamal Anadkat, et~al.
\newblock Gpt-4 technical report.
\newblock {\em arXiv preprint arXiv:2303.08774}, 2023.

\bibitem{lacoste2019quantifying}
Alexandre Lacoste, Alexandra Luccioni, Victor Schmidt, and Thomas Dandres.
\newblock Quantifying the carbon emissions of machine learning.
\newblock {\em arXiv preprint arXiv:1910.09700}, 2019.

\bibitem{lannelongue2021green}
Lo{\"\i}c Lannelongue, Jason Grealey, and Michael Inouye.
\newblock Green algorithms: quantifying the carbon footprint of computation.
\newblock {\em Advanced science}, 8(12):2100707, 2021.

\bibitem{JMLR:v24:23-0069}
Alexandra~Sasha Luccioni, Sylvain Viguier, and Anne-Laure Ligozat.
\newblock Estimating the carbon footprint of bloom, a 176b parameter language model.
\newblock {\em Journal of Machine Learning Research}, 24(253):1--15, 2023.

\bibitem{workshop2022bloom}
BigScience Workshop, Teven~Le Scao, Angela Fan, Christopher Akiki, Ellie Pavlick, Suzana Ili{\'c}, Daniel Hesslow, Roman Castagn{\'e}, Alexandra~Sasha Luccioni, Fran{\c{c}}ois Yvon, et~al.
\newblock Bloom: A 176b-parameter open-access multilingual language model.
\newblock {\em arXiv preprint arXiv:2211.05100}, 2022.

\bibitem{faiz2023llmcarbon}
Ahmad Faiz, Sotaro Kaneda, Ruhan Wang, Rita~Chukwunyere Osi, Prateek Sharma, Fan Chen, and Lei Jiang.
\newblock Llmcarbon: Modeling the end-to-end carbon footprint of large language models.
\newblock In {\em The Twelfth International Conference on Learning Representations}, 2023.

\bibitem{little2008little}
John~DC Little and Stephen~C Graves.
\newblock Little's law.
\newblock {\em Building intuition: insights from basic operations management models and principles}, pages 81--100, 2008.

\bibitem{narayanan2021efficient}
Deepak Narayanan, Mohammad Shoeybi, Jared Casper, Patrick LeGresley, Mostofa Patwary, Vijay Korthikanti, Dmitri Vainbrand, Prethvi Kashinkunti, Julie Bernauer, Bryan Catanzaro, et~al.
\newblock Efficient large-scale language model training on gpu clusters using megatron-lm.
\newblock In {\em Proceedings of the International Conference for High Performance Computing, Networking, Storage and Analysis}, pages 1--15, 2021.

\bibitem{epoch2023aitrends}
{Epoch AI}.
\newblock Key trends and figures in machine learning, 2023.
\newblock Accessed: 2024-05-20.

\bibitem{zeng2022glm}
Aohan Zeng, Xiao Liu, Zhengxiao Du, Zihan Wang, Hanyu Lai, Ming Ding, Zhuoyi Yang, Yifan Xu, Wendi Zheng, Xiao Xia, et~al.
\newblock Glm-130b: An open bilingual pre-trained model.
\newblock In {\em The Eleventh International Conference on Learning Representations}, 2022.

\bibitem{li2023starcoder}
Raymond Li, Yangtian Zi, Niklas Muennighoff, Denis Kocetkov, Chenghao Mou, Marc Marone, Christopher Akiki, LI~Jia, Jenny Chim, Qian Liu, et~al.
\newblock Starcoder: may the source be with you!
\newblock {\em Transactions on Machine Learning Research}, 2023.

\bibitem{dosovitskiy2020image}
Alexey Dosovitskiy, Lucas Beyer, Alexander Kolesnikov, Dirk Weissenborn, Xiaohua Zhai, Thomas Unterthiner, Mostafa Dehghani, Matthias Minderer, Georg Heigold, Sylvain Gelly, et~al.
\newblock An image is worth 16x16 words: Transformers for image recognition at scale.
\newblock In {\em International Conference on Learning Representations}, 2020.

\bibitem{liu2021swin}
Ze~Liu, Yutong Lin, Yue Cao, Han Hu, Yixuan Wei, Zheng Zhang, Stephen Lin, and Baining Guo.
\newblock Swin transformer: Hierarchical vision transformer using shifted windows.
\newblock In {\em Proceedings of the IEEE/CVF international conference on computer vision}, pages 10012--10022, 2021.

\bibitem{hendryckstest2021}
Dan Hendrycks, Collin Burns, Steven Basart, Andy Zou, Mantas Mazeika, Dawn Song, and Jacob Steinhardt.
\newblock Measuring massive multitask language understanding.
\newblock {\em Proceedings of the International Conference on Learning Representations (ICLR)}, 2021.

\bibitem{deng2009imagenet}
Jia Deng, Wei Dong, Richard Socher, Li-Jia Li, Kai Li, and Li~Fei-Fei.
\newblock Imagenet: A large-scale hierarchical image database.
\newblock In {\em 2009 IEEE conference on computer vision and pattern recognition}, pages 248--255. Ieee, 2009.

\bibitem{wu2022sustainable}
Carole-Jean Wu, Ramya Raghavendra, Udit Gupta, Bilge Acun, Newsha Ardalani, Kiwan Maeng, Gloria Chang, Fiona Aga, Jinshi Huang, Charles Bai, et~al.
\newblock Sustainable ai: Environmental implications, challenges and opportunities.
\newblock {\em Proceedings of Machine Learning and Systems}, 4:795--813, 2022.

\bibitem{trebaol2020cumulator}
Tristan Trébaol.
\newblock Cumulator—a tool to quantify and report the carbon footprint of machine learning computations and communication in academia and healthcare.
\newblock {\em Infoscience EPFL: record 278189}, 2020.

\bibitem{lottick2019energy}
Kadan Lottick, Silvia Susai, Sorelle~A Friedler, and Jonathan~P Wilson.
\newblock Energy usage reports: Environmental awareness as part of algorithmic accountability.
\newblock {\em arXiv preprint arXiv:1911.08354}, 2019.

\bibitem{budennyy2022eco2ai}
Semen~Andreevich Budennyy, Vladimir~Dmitrievich Lazarev, Nikita~Nikolaevich Zakharenko, Aleksei~N Korovin, OA~Plosskaya, Denis~Valer’evich Dimitrov, VS~Akhripkin, IV~Pavlov, Ivan~Valer'evich Oseledets, Ivan~Segundovich Barsola, et~al.
\newblock Eco2ai: carbon emissions tracking of machine learning models as the first step towards sustainable ai.
\newblock In {\em Doklady Mathematics}, volume 106, pages S118--S128. Springer, 2022.

\bibitem{anthony2020carbontracker}
Lasse F~Wolff Anthony, Benjamin Kanding, and Raghavendra Selvan.
\newblock Carbontracker: Tracking and predicting the carbon footprint of training deep learning models.
\newblock {\em arXiv preprint arXiv:2007.03051}, 2020.

\bibitem{henderson2020towards}
Peter Henderson, Jieru Hu, Joshua Romoff, Emma Brunskill, Dan Jurafsky, and Joelle Pineau.
\newblock Towards the systematic reporting of the energy and carbon footprints of machine learning.
\newblock {\em Journal of Machine Learning Research}, 21(248):1--43, 2020.

\bibitem{team2023gemini}
Gemini Team, Rohan Anil, Sebastian Borgeaud, Yonghui Wu, Jean-Baptiste Alayrac, Jiahui Yu, Radu Soricut, Johan Schalkwyk, Andrew~M Dai, Anja Hauth, et~al.
\newblock Gemini: a family of highly capable multimodal models.
\newblock {\em arXiv preprint arXiv:2312.11805}, 2023.

\bibitem{jiang2024megascale}
Ziheng Jiang, Haibin Lin, Yinmin Zhong, Qi~Huang, Yangrui Chen, Zhi Zhang, Yanghua Peng, Xiang Li, Cong Xie, Shibiao Nong, et~al.
\newblock Megascale: Scaling large language model training to more than 10,000 gpus.
\newblock {\em arXiv preprint arXiv:2402.15627}, 2024.

\bibitem{openai2023gpt4}
OpenAI.
\newblock Gpt-4 technical report, 2023.

\bibitem{anil2023palm}
Rohan Anil, Andrew~M Dai, Orhan Firat, Melvin Johnson, Dmitry Lepikhin, Alexandre Passos, Siamak Shakeri, Emanuel Taropa, Paige Bailey, Zhifeng Chen, et~al.
\newblock Palm 2 technical report.
\newblock {\em arXiv preprint arXiv:2305.10403}, 2023.

\bibitem{chung2024scaling}
Hyung~Won Chung, Le~Hou, Shayne Longpre, Barret Zoph, Yi~Tay, William Fedus, Yunxuan Li, Xuezhi Wang, Mostafa Dehghani, Siddhartha Brahma, et~al.
\newblock Scaling instruction-finetuned language models.
\newblock {\em Journal of Machine Learning Research}, 25(70):1--53, 2024.

\bibitem{thoppilan2022lamda}
Romal Thoppilan, Daniel De~Freitas, Jamie Hall, Noam Shazeer, Apoorv Kulshreshtha, Heng-Tze Cheng, Alicia Jin, Taylor Bos, Leslie Baker, Yu~Du, et~al.
\newblock Lamda: Language models for dialog applications.
\newblock {\em arXiv preprint arXiv:2201.08239}, 2022.

\bibitem{touvron2023llama}
Hugo Touvron, Louis Martin, Kevin Stone, Peter Albert, Amjad Almahairi, Yasmine Babaei, Nikolay Bashlykov, Soumya Batra, Prajjwal Bhargava, Shruti Bhosale, et~al.
\newblock Llama 2: Open foundation and fine-tuned chat models.
\newblock {\em arXiv preprint arXiv:2307.09288}, 2023.

\bibitem{wei2023skywork}
Tianwen Wei, Liang Zhao, Lichang Zhang, Bo~Zhu, Lijie Wang, Haihua Yang, Biye Li, Cheng Cheng, Weiwei L{\"u}, Rui Hu, et~al.
\newblock Skywork: A more open bilingual foundation model.
\newblock {\em arXiv preprint arXiv:2310.19341}, 2023.

\bibitem{wu2023bloomberggpt}
Shijie Wu, Ozan Irsoy, Steven Lu, Vadim Dabravolski, Mark Dredze, Sebastian Gehrmann, Prabhanjan Kambadur, David Rosenberg, and Gideon Mann.
\newblock Bloomberggpt: A large language model for finance.
\newblock {\em arXiv preprint arXiv:2303.17564}, 2023.

\bibitem{9477085}
Ahmed Elnaggar, Michael Heinzinger, Christian Dallago, Ghalia Rehawi, Wang Yu, Llion Jones, Tom Gibbs, Tamas Feher, Christoph Angerer, Martin Steinegger, Debsindhu Bhowmik, and Burkhard Rost.
\newblock Prottrans: Towards cracking the language of lifes code through self-supervised deep learning and high performance computing.
\newblock {\em IEEE Transactions on Pattern Analysis and Machine Intelligence}, pages 1--1, 2021.

\bibitem{esm2}
Zeming Lin, Halil Akin, Roshan Rao, Brian Hie, Zhongkai Zhu, Wenting Lu, Nikita Smetanin, Robert Verkuil, Ori Kabeli, Yaniv Shmueli, et~al.
\newblock Evolutionary-scale prediction of atomic-level protein structure with a language model.
\newblock {\em Science}, 379(6637):1123--1130, 2023.

\bibitem{xTrimoPGLM}
Bo~Chen, Xingyi Cheng, Pan Li, Yangli-ao Geng, Jing Gong, Shen Li, Zhilei Bei, Xu~Tan, Boyan Wang, Xin Zeng, et~al.
\newblock xtrimopglm: unified 100b-scale pre-trained transformer for deciphering the language of protein.
\newblock {\em arXiv preprint arXiv:2401.06199}, 2024.

\bibitem{bi2022pangu}
Kaifeng Bi, Lingxi Xie, Hengheng Zhang, Xin Chen, Xiaotao Gu, and Qi~Tian.
\newblock Pangu-weather: A 3d high-resolution model for fast and accurate global weather forecast.
\newblock {\em arXiv preprint arXiv:2211.02556}, 2022.

\bibitem{ren2023pangu}
Xiaozhe Ren, Pingyi Zhou, Xinfan Meng, Xinjing Huang, Yadao Wang, Weichao Wang, Pengfei Li, Xiaoda Zhang, Alexander Podolskiy, Grigory Arshinov, et~al.
\newblock Pangu-$\{$$\backslash$Sigma$\}$: Towards trillion parameter language model with sparse heterogeneous computing.
\newblock {\em arXiv preprint arXiv:2303.10845}, 2023.

\bibitem{vit22b}
Mostafa Dehghani, Josip Djolonga, Basil Mustafa, Piotr Padlewski, Jonathan Heek, Justin Gilmer, Andreas~Peter Steiner, Mathilde Caron, Robert Geirhos, Ibrahim Alabdulmohsin, et~al.
\newblock Scaling vision transformers to 22 billion parameters.
\newblock In {\em International Conference on Machine Learning}, pages 7480--7512. PMLR, 2023.

\bibitem{sdldm}
Robin Rombach, Andreas Blattmann, Dominik Lorenz, Patrick Esser, and Bj{\"o}rn Ommer.
\newblock High-resolution image synthesis with latent diffusion models.
\newblock In {\em Proceedings of the IEEE/CVF conference on computer vision and pattern recognition}, pages 10684--10695, 2022.

\bibitem{taiyi}
Jiaxing Zhang, Ruyi Gan, Junjie Wang, Yuxiang Zhang, Lin Zhang, Ping Yang, Xinyu Gao, Ziwei Wu, Xiaoqun Dong, Junqing He, Jianheng Zhuo, Qi~Yang, Yongfeng Huang, Xiayu Li, Yanghan Wu, Junyu Lu, Xinyu Zhu, Weifeng Chen, Ting Han, Kunhao Pan, Rui Wang, Hao Wang, Xiaojun Wu, Zhongshen Zeng, and Chongpei Chen.
\newblock Fengshenbang 1.0: Being the foundation of chinese cognitive intelligence.
\newblock {\em CoRR}, abs/2209.02970, 2022.

\bibitem{gshard}
Dmitry Lepikhin, HyoukJoong Lee, Yuanzhong Xu, Dehao Chen, Orhan Firat, Yanping Huang, Maxim Krikun, Noam Shazeer, and Zhifeng Chen.
\newblock Gshard: Scaling giant models with conditional computation and automatic sharding.
\newblock {\em arXiv preprint arXiv:2006.16668}, 2020.

\bibitem{nllb}
Marta~R Costa-juss{\`a}, James Cross, Onur {\c{C}}elebi, Maha Elbayad, Kenneth Heafield, Kevin Heffernan, Elahe Kalbassi, Janice Lam, Daniel Licht, Jean Maillard, et~al.
\newblock No language left behind: Scaling human-centered machine translation.
\newblock {\em arXiv preprint arXiv:2207.04672}, 2022.

\bibitem{imagen}
Chitwan Saharia, William Chan, Saurabh Saxena, Lala Li, Jay Whang, Emily~L Denton, Kamyar Ghasemipour, Raphael Gontijo~Lopes, Burcu Karagol~Ayan, Tim Salimans, et~al.
\newblock Photorealistic text-to-image diffusion models with deep language understanding.
\newblock {\em Advances in neural information processing systems}, 35:36479--36494, 2022.

\bibitem{parti}
Jiahui Yu, Yuanzhong Xu, Jing~Yu Koh, Thang Luong, Gunjan Baid, Zirui Wang, Vijay Vasudevan, Alexander Ku, Yinfei Yang, Burcu~Karagol Ayan, et~al.
\newblock Scaling autoregressive models for content-rich text-to-image generation.
\newblock {\em arXiv preprint arXiv:2206.10789}, 2(3):5, 2022.

\bibitem{flamingo}
Jean-Baptiste Alayrac, Jeff Donahue, Pauline Luc, Antoine Miech, Iain Barr, Yana Hasson, Karel Lenc, Arthur Mensch, Katherine Millican, Malcolm Reynolds, et~al.
\newblock Flamingo: a visual language model for few-shot learning.
\newblock {\em Advances in neural information processing systems}, 35:23716--23736, 2022.

\bibitem{pseudolabel}
Hieu Pham, Zihang Dai, Qizhe Xie, and Quoc~V Le.
\newblock Meta pseudo labels.
\newblock In {\em Proceedings of the IEEE/CVF conference on computer vision and pattern recognition}, pages 11557--11568, 2021.

\bibitem{coatnet}
Zihang Dai, Hanxiao Liu, Quoc~V Le, and Mingxing Tan.
\newblock Coatnet: Marrying convolution and attention for all data sizes.
\newblock {\em Advances in neural information processing systems}, 34:3965--3977, 2021.

\bibitem{coca}
Jiahui Yu, Zirui Wang, Vijay Vasudevan, Legg Yeung, Mojtaba Seyedhosseini, and Yonghui Wu.
\newblock Coca: Contrastive captioners are image-text foundation models.
\newblock {\em arXiv preprint arXiv:2205.01917}, 2022.

\bibitem{basicl}
Hieu Pham, Zihang Dai, Golnaz Ghiasi, Kenji Kawaguchi, Hanxiao Liu, Adams~Wei Yu, Jiahui Yu, Yi-Ting Chen, Minh-Thang Luong, Yonghui Wu, et~al.
\newblock Combined scaling for zero-shot transfer learning.
\newblock {\em Neurocomputing}, 555:126658, 2023.

\bibitem{rajbhandari2020zero}
Samyam Rajbhandari, Jeff Rasley, Olatunji Ruwase, and Yuxiong He.
\newblock Zero: Memory optimizations toward training trillion parameter models.
\newblock In {\em SC20: International Conference for High Performance Computing, Networking, Storage and Analysis}, pages 1--16. IEEE, 2020.

\bibitem{adiwardana2020towards}
Daniel Adiwardana, Minh-Thang Luong, David~R So, Jamie Hall, Noah Fiedel, Romal Thoppilan, Zi~Yang, Apoorv Kulshreshtha, Gaurav Nemade, Yifeng Lu, et~al.
\newblock Towards a human-like open-domain chatbot.
\newblock {\em arXiv preprint arXiv:2001.09977}, 2020.

\bibitem{fedus2022switch}
William Fedus, Barret Zoph, and Noam Shazeer.
\newblock Switch transformers: Scaling to trillion parameter models with simple and efficient sparsity.
\newblock {\em Journal of Machine Learning Research}, 23(120):1--39, 2022.

\bibitem{clip}
Alec Radford, Jong~Wook Kim, Chris Hallacy, Aditya Ramesh, Gabriel Goh, Sandhini Agarwal, Girish Sastry, Amanda Askell, Pamela Mishkin, Jack Clark, et~al.
\newblock Learning transferable visual models from natural language supervision.
\newblock In {\em International conference on machine learning}, pages 8748--8763. PMLR, 2021.

\bibitem{igpt}
Mark Chen, Alec Radford, Rewon Child, Jeffrey Wu, Heewoo Jun, David Luan, and Ilya Sutskever.
\newblock Generative pretraining from pixels.
\newblock In {\em International conference on machine learning}, pages 1691--1703. PMLR, 2020.

\bibitem{alphastar}
Oriol Vinyals, Igor Babuschkin, Wojciech~M Czarnecki, Micha{\"e}l Mathieu, Andrew Dudzik, Junyoung Chung, David~H Choi, Richard Powell, Timo Ewalds, Petko Georgiev, et~al.
\newblock Grandmaster level in starcraft ii using multi-agent reinforcement learning.
\newblock {\em Nature}, 575(7782):350--354, 2019.

\bibitem{bi2024deepseek}
Xiao Bi, Deli Chen, Guanting Chen, Shanhuang Chen, Damai Dai, Chengqi Deng, Honghui Ding, Kai Dong, Qiushi Du, Zhe Fu, et~al.
\newblock Deepseek llm: Scaling open-source language models with longtermism.
\newblock {\em arXiv preprint arXiv:2401.02954}, 2024.

\bibitem{young2024yi}
Alex Young, Bei Chen, Chao Li, Chengen Huang, Ge~Zhang, Guanwei Zhang, Heng Li, Jiangcheng Zhu, Jianqun Chen, Jing Chang, et~al.
\newblock Yi: Open foundation models by 01. ai.
\newblock {\em arXiv preprint arXiv:2403.04652}, 2024.

\bibitem{swinv2}
Ze~Liu, Han Hu, Yutong Lin, Zhuliang Yao, Zhenda Xie, Yixuan Wei, Jia Ning, Yue Cao, Zheng Zhang, Li~Dong, et~al.
\newblock Swin transformer v2: Scaling up capacity and resolution.
\newblock In {\em Proceedings of the IEEE/CVF conference on computer vision and pattern recognition}, pages 12009--12019, 2022.

\bibitem{vitg14}
Xiaohua Zhai, Alexander Kolesnikov, Neil Houlsby, and Lucas Beyer.
\newblock Scaling vision transformers.
\newblock In {\em Proceedings of the IEEE/CVF conference on computer vision and pattern recognition}, pages 12104--12113, 2022.

\bibitem{eva01}
Yuxin Fang, Wen Wang, Binhui Xie, Quan Sun, Ledell Wu, Xinggang Wang, Tiejun Huang, Xinlong Wang, and Yue Cao.
\newblock Eva: Exploring the limits of masked visual representation learning at scale.
\newblock In {\em Proceedings of the IEEE/CVF Conference on Computer Vision and Pattern Recognition}, pages 19358--19369, 2023.

\bibitem{noisystudent}
Qizhe Xie, Minh-Thang Luong, Eduard Hovy, and Quoc~V Le.
\newblock Self-training with noisy student improves imagenet classification.
\newblock In {\em Proceedings of the IEEE/CVF conference on computer vision and pattern recognition}, pages 10687--10698, 2020.

\bibitem{align}
Chao Jia, Yinfei Yang, Ye~Xia, Yi-Ting Chen, Zarana Parekh, Hieu Pham, Quoc Le, Yun-Hsuan Sung, Zhen Li, and Tom Duerig.
\newblock Scaling up visual and vision-language representation learning with noisy text supervision.
\newblock In {\em International conference on machine learning}, pages 4904--4916. PMLR, 2021.

\bibitem{seer}
Priya Goyal, Mathilde Caron, Benjamin Lefaudeux, Min Xu, Pengchao Wang, Vivek Pai, Mannat Singh, Vitaliy Liptchinsky, Ishan Misra, Armand Joulin, et~al.
\newblock Self-supervised pretraining of visual features in the wild.
\newblock {\em arXiv preprint arXiv:2103.01988}, 2021.

\bibitem{resnetrs}
Irwan Bello, William Fedus, Xianzhi Du, Ekin~Dogus Cubuk, Aravind Srinivas, Tsung-Yi Lin, Jonathon Shlens, and Barret Zoph.
\newblock Revisiting resnets: Improved training and scaling strategies.
\newblock {\em Advances in Neural Information Processing Systems}, 34:22614--22627, 2021.

\end{thebibliography}
}






\newpage
\appendix

\section*{Appendix}

\section{Additional information of the evaluated model}
The information of evaluated models~\cref{fig:timeline,fig:trade} are mostly from EpochAI~\cite{epoch2023aitrends} and the carbon intensity of different regions is from Electricity Maps~\footnote{https://app.electricitymaps.com/map}.
The MMLU~\cite{hendryckstest2021} and ImageNet-1k Accuracy~\cite{deng2009imagenet} performances are collected from their published results and PapersWithCode~\footnote{https://paperswithcode.com/sota/image-classification-on-imagenet}.

Figure 1 provides a comprehensive overview of the carbon footprint of large-scale AI models, spanning 42 models across 15 tasks, as systematically classified by EpochAI~\cite{epoch2023aitrends}.

\textbf{Chat} LLaMa-3-70B~\cite{llama3modelcard}, Inflection 2.5~\footnote{https://inflection.ai/}.

\textbf{Language model} Gemini Ultra~\cite{team2023gemini}, MegaScale (Prduction)~\cite{jiang2024megascale}, Inflection 2
, GPT-4~\cite{openai2023gpt4}, PaLM-2~\cite{anil2023palm}, GPT-3.5, Flan-PaLM 540B, Flan-T5-11B, Flan-137B~\cite{chung2024scaling}, Megatron-Turing NLG 530B~\cite{narayanan2021efficient}, LaMDA~\cite{thoppilan2022lamda}, LLaMa~\cite{touvron2023llama},LLaMa-2~\cite{touvron2023llama}, BLOOM~\cite{workshop2022bloom}, Skywork-13B~\cite{wei2023skywork}, BloombergGPT~\cite{wu2023bloomberggpt}.

\textbf{Proteins}
ProT5-XXL~\cite{9477085}, ESM2-15B~\cite{esm2}, xTrimoPGLM -100B~\cite{xTrimoPGLM}.

\textbf{Weather prediction}
Pangu Weather~\cite{bi2022pangu}.

\textbf{Code generation}
Pangu-$\Sigma$~\cite{ren2023pangu}, StarCoder~\cite{li2023starcoder}.

\textbf{Object detection}
ViT-22B ~\cite{vit22b}

\textbf{Image generation}
Stable Diffusion (LDM-KL-8-G) ~\cite{sdldm}, Taiyi-Stable Diffusion~\cite{taiyi}

\textbf{Translation}
Gshard (dense)~\cite{gshard}, NLLB~\cite{nllb}

\textbf{Text-to-image}
Imagen~\cite{imagen}, Parti~\cite{parti}.

\textbf{Visual question answering}
Flamingo~\cite{flamingo}.

\textbf{Image classification}
Meta Pseudo Label ~\cite{pseudolabel}, CoAtNet~\cite{coatnet}, CoCa~\cite{coca}, BASIC-L~\cite{basicl}.

\textbf{Text autocompletion}
GPT-3-175B~\cite{brown2020language}, Turing-NLG~\cite{rajbhandari2020zero}, Meena~\cite{adiwardana2020towards}, Switch~\cite{fedus2022switch}.

\textbf{Zero-shot image classification}
CLIP (ViT L/14@336px)~\cite{clip}.

\textbf{Image completion}
iGPT-XL~\cite{igpt}.

\textbf{StarCraft}
AlphaStar~\cite{alphastar}.

The language models featured in \cref{fig:trade}, includes BLOOM~\cite{workshop2022bloom}, StarCoder~\cite{li2023starcoder}, GLM~\cite{zeng2022glm}, GPT-3~\cite{brown2020language}, DeepseekLLM-67B~\cite{bi2024deepseek}, Yi-34B~\cite{young2024yi}, and LLaMa-3-70B~\cite{llama3modelcard}.

In addition, the vision models showcased in \cref{fig:trade}, include CoCa~\cite{coca}, BASIC-L~\cite{basicl}, CoAtNet~\cite{coatnet}, Swin Transformer V2~\cite{swinv2}, ViT-G/14~\cite{vitg14}, Meta Pseudo Label~\cite{pseudolabel}, EVA-01~\cite{eva01}, Noisy Student (L2)~\cite{noisystudent}, ALIGN~\cite{align}, SEER~\cite{seer}, ResNet-RS~\cite{resnetrs}, were utilized to assess their carbon footprint in relation to the Top-1 Accuracy of image classification on the ImageNet-1K~\cite{deng2009imagenet} benchmark.

\end{document}